# Instability of Taylor-Couette Flow between Concentric Rotating Cylinders

Hua-Shu Dou[1*], Boo Cheong Khoo[2], and Khoon Seng Yeo[2]

[1]Temasek Laboratories,
National University of Singapore, Singapore 117508
[2]Department of Mechanical Engineering,
National University of Singapore, Singapore 119260

*Email: tsldh@nus.edu.sg; huashudou@yahoo.com

**Abstract** The energy gradient theory is used to study the instability of Taylor-Couette flow between concentric rotating cylinders. This theory has been proposed in our previous works. In our previous studies, the energy gradient theory was demonstrated to be applicable for wall-bounded parallel flows. It was found that the critical value of the energy gradient parameter $K_{max}$ at turbulent transition is about 370-389 for wall-bounded parallel flows (which include plane Poiseuille flow, pipe Poiseuille flow and plane Couette flow) below which no turbulence occurs. In this paper, the detailed derivation for the calculation of the energy gradient parameter in the flow between concentric rotating cylinders is provided. The calculated results for the critical condition of primary instability (with semi-empirical treatment) are found to be in very good agreement with the experiments in the literature. A possible mechanism of spiral turbulence generation observed for counter-rotation of two cylinders can also be explained using the energy gradient theory. The energy gradient theory can serve to relate the condition of transition in Taylor-Couette flow to that in plane Couette flow. The latter reasonably becomes the limiting case of the former when the radii of cylinders tend to infinity. It is our contention that the energy gradient theory is possibly fairly universal for analysis of flow instability and turbulent transition, and is found valid for both pressure and shear driven flows in parallel and rotating flow configurations.

**Nomenclature**

A, $A_a$, A*    coefficients    $s^{-1}$

$\overline{A}$    amplitude of the disturbance distance    m

B, $B_a$, B*    coefficients    $m^2 s^{-1}$

D    diameter of the pipe for pipe flow    m

E    total mechanical energy of unit volume of fluid    $J m^{-3}$

h    $= R_2 - R_1$, gap width between the inner cylinder and the outer cylinder    m

H    total mechanical energy loss of unit volume of fluid due to viscosity in streamwise direction    $J m^{-3}$

K    function of coordinates (dimensionless).

$K_c$    critical value of $K_{max}$ for instability (dimensionless).

$K_{max}$    maximum of K in the domain (dimensionless).

l    half-width of the channel for plane Poiseuille flow and plane Couette flow    m

n    coordinate in transverse direction    m

p    static pressure    $N m^{-2}$

r    radius    m

$R_0$    average radius of inner cylinder and outer cylinder    m

$R_1$    radius of inner cylinder    m

$R_2$    radius of outer cylinder    m

Re    Reynolds number (dimensionless).

s    coordinate in streamwise direction    m

t    time    s

T    Taylor number (dimensionless).

u    velocity component in the main flow direction    $m s^{-1}$

$u_0$    velocity at the mid-plane for plane Poiseuille flow (channel flow)    $m s^{-1}$

U    average velocity in the flow passage    $m s^{-1}$

v    velocity component in the transverse direction    $m s^{-1}$

$v'_m$    $= \overline{A} \omega_d$, amplitude of the disturbance of velocity in transverse direction    $m s^{-1}$

W    work done to the unit volumetric fluid by external    $J m^{-3}$

x    coordinate in the streamwise direction    m

y    coordinate in the transverse direction    m

z    coordinate in the spanwise direction    m

| | | |
|---|---|---|
| η | radius ratio, $\equiv R_2/R_1$ | |
| θ | angular coordinates | rad |
| λ | speed ratio, $\equiv \omega_2/\omega_1$ | |
| μ | dynamics viscosity | Nm$^{-2}$ s |
| ν | kinematic viscosity | m²s$^{-1}$ |
| ρ | density of fluid | kg m$^{-3}$ |
| τ | shear stress | N m$^{-2}$ |
| ω | angular velocity of the fluid | rad s$^{-1}$ |
| $\omega_1$ | angular velocity of the inner cylinder | rad s$^{-1}$ |
| $\omega_2$ | angular velocity of the outer cylinder | rad s$^{-1}$ |
| $\omega_{1a}$ | angular velocity of the inner cylinder after splitting | rad s$^{-1}$ |
| $\omega_{2a}$ | angular velocity of the outer cylinder after splitting | rad s$^{-1}$ |
| $\omega_d$ | frequency of the disturbance | s$^{-1}$ |

## 1. Introduction

Taylor-Couette flow refers to the problem of flow between two concentric rotating cylinders as shown in Fig.1 [1-4]. This terminology was named after the works of G. I. Taylor (1923) and M. Couette (1890). This problem was first investigated experimentally by Couette (1890) and Mallock (1896). Couette observed that the torque needed to rotate the outer cylinder increased linearly with the rotation speed until a critical rotation speed, after which the torque increased more rapidly. This change was due to a transition from stable to unstable flow at the critical rotation speed. Taylor was the first to successfully apply linear stability theory to a specific problem, and succeeded in obtaining an excellent agreement of theory with experiments for the flow instability between two concentric rotating cylinders [5]. Taylor's groundbreaking research for this problem has been considered as a classical example of flow instability study [6-8].

In the past years, the problem of Taylor-Couette flow has received renewed interests because of its importance in flow stability and the fact that it is particularly amenable to rigorous mathematical treatment/analysis due to infinitesimal disturbances [1-3]. For the stability of an inviscid fluid moving in concentric layers, Lord Rayleigh [9] used the circulation variation versus the radius to explain the instability while von Karman [10] employed the relative roles of centrifugal force and pressure gradient to interpret the instability initiation. Their goal was to determine the condition for which a perturbation resulting from an adverse gradient of angular momentum can be unstable. In his classic paper, Taylor [5] presented a mathematical stability analysis for viscous flow and compared the results to laboratory observations. Taylor observed that, for small ratio of the gap width to the cylinder radii and for a given rotating speed of outer cylinder, when the rotation speed of the inner cylinder is low, the flow remains laminar; when the rotation speed of the inner cylinder exceeds a critical value, instability sets in and rows of cellular vortices are developed. When the rotating speed is increased to an even higher value, the cell rows break down and a turbulence pattern is produced. He proposed a parameter, now commonly known as the Taylor number, $T = \text{Re}^2 (h/R_0)$, to characterize this critical condition for instability. Here, Re is the Reynolds number based on the gap width (h) and the rotation speed of the inner cylinder, and $R_0$ is the mean radius of the inner cylinder and the outer cylinder. The critical value of the Taylor number for primary instability is 1708 as obtained from linear analysis. This value agrees well with his experiments [1-3]. For Taylor-Couette flow, Snyder has given a semi-empirical equation for the critical condition from the collected experimental data [11]. Esser and Grossmann have also given an analytical equation for the critical condition by an simple approximation, but a constant in the equation have to be fixed using the result of linear stability analysis [12].

However, the problem of Taylor-Couette flow is still far from completely resolved despite extensive study [11-17]. For example, the limiting case of Taylor-Couette flow when the ratio of the gap width to the radii tends to zero should agree with that of plane Couette flow. This includes two possibilities: either radius is infinite or gap width is very small. Thus, the criterion for instability should reflect this phenomenon. There are some recent works trying to address this issue to some degree of success [18-20]. One may observes that Taylor's criterion is not appropriate when this limiting case is studied because plane Couette flow is judged to be always stable due to Taylor number assuming a null value using Taylor's criterion. This may be attributed to the fact that Taylor's criterion only considered the effect of centrifugal force, and does not include the kinematic inertia force. Therefore, it is reckoned to be suitable for low Re number flows with high curvature. For rotating flow with higher Re number and low curvature, the flow may transit to turbulence earlier and yet does not violate Taylor's criterion.

Recently, Dou [21,22] proposed a new energy gradient theory to analyze flow instability and turbulent transition problems. In this theory, the critical condition for flow instability depends both on the base flow and the disturbance which agrees with the experimental observations. For a given disturbance, the critical condition for flow instability and turbulent transition is determined by the ratio (K) of the gradient of total mechanical energy in the transverse direction to the loss of total mechanical energy in the streamwise direction. For a given flow geometry and fluid properties, when the maximum of *K* in the flow field is larger than a critical value, it is expected that instability would occur for some initial disturbances provided that the disturbance energy is sufficiently large. For plane Poiseuille flow (channel flow), Hagen-Poiseuille flow (pipe flow), and plane Couette flow (simple shear flow), the findings based on the theory are consistent with the experimental observations; for the experimental determined critical condition, $K_c$=370-389 for all the above mentioned three types of flows below which there is no occurrence of turbulence. In these comparisons, the distribution of K was calculated for each flow and the value of $K_c$ was obtained using the experimental data at critical condition [21, 22, 23]. The theory also suggests the mechanism of instability associated with an inflectional velocity profile for viscous flows. The theory has been extended to curved flows with similar derivations to parallel flows and three important theorems have been obtained [24]. This theory has also been employed to study the viscoelastic flows where the effect of elastic force is dominating [25]. It should be mentioned that the energy gradient theory is a semi-empirical theory since the critical value of K is observed and determined experimentally and can not be directly calculated from the theory so far. In this theory, only the critical condition for the instability is sought after and the detailed process of instability is not provided.

In this study, we apply the energy gradient theory to analyze the Taylor-Couette flow between concentric rotating cylinders, and aim to demonstrate that the mechanism of instability in Taylor-Couette flow can be explained via the energy gradient concept. Through comparison with experiments, we show that the energy gradient function K as a stability criterion is sufficient to describe and characterize the flow instability in Taylor-Couette flow. We also show that plane Couette flow can be considered as just the limiting case of Taylor-Couette flow when the curvature of the walls tends to zero. For flow between concentric rotating cylinders, the flow instability may be induced by rotation of the inner cylinder or the outer cylinder. If it is induced by the former, a Taylor vortex cell pattern will be formed when the critical condition is violated as in the experiments; if it is induced by the latter, Taylor vortex cell pattern will not occur and the flow may directly transit to turbulence when the critical condition due to inertia force is reached as in plane Couette flow [1-3, 6]. In this study, only the critical condition for the former situation is considered/treated.

**2. Energy gradient theory revisted**

Dou [21] proposed a mechanism with the aim to clarify the phenomenon of transition from laminar flow to turbulence for wall-bounded shear flows. In this mechanism, the whole flow field is treated as an energy field. It is proposed that the gradient of total mechanical energy in the transverse direction of the main flow and the total mechanical energy loss from viscous friction in the streamwise direction dominate the instability phenomena and hence the flow transition for a given disturbance. It is suggested that the energy gradient in the transverse direction has the potential to amplify a velocity disturbance, while the viscous friction loss in the streamwise direction can resist and absorb this disturbance. The flow instability or the transition to turbulence depends on the relative magnitude of these two roles of energy gradient amplification and viscous friction damping of the initial disturbance. In [22], more detailed derivation has been given to exactly describe this mechanism, and this theory is termed as "energy gradient theory." Here, we give a short discussion for a better understanding of the work presented in this study.

The equation of total mechanical energy for incompressible flow by neglecting the gravitational energy can be written as [21],

$$\rho \frac{\partial \mathbf{u}}{\partial t} + \nabla(p + \frac{1}{2}\rho u^2) = \mu \nabla^2 \mathbf{u} + \rho(\mathbf{u} \times \nabla \times \mathbf{u}). \qquad (1)$$

For pressure driven flows, the derivatives of the total mechanical energy in the transverse direction and the streamwise direction can be expressed, respectively, as [21-24],

$$\frac{\partial E}{\partial n} = \frac{\partial(p + (1/2)\rho u^2)}{\partial n} = \rho(\mathbf{u} \times \boldsymbol{\omega}) \cdot \frac{d\mathbf{n}}{|d\mathbf{n}|} + (\mu \nabla^2 \mathbf{u}) \cdot \frac{d\mathbf{n}}{|d\mathbf{n}|} = \rho u \omega + (\mu \nabla^2 \mathbf{u})_n, \qquad (2)$$

$$\frac{\partial E}{\partial s} = \frac{\partial(p + (1/2)\rho u^2)}{\partial s} = \rho(\mathbf{u} \times \boldsymbol{\omega}) \cdot \frac{d\mathbf{s}}{|d\mathbf{s}|} + (\mu \nabla^2 \mathbf{u}) \cdot \frac{d\mathbf{s}}{|d\mathbf{s}|} = (\mu \nabla^2 \mathbf{u})_s, \qquad (3)$$

where $\boldsymbol{\omega} = \nabla \times \mathbf{u}$ is the vorticity. Since there is no work input in the pressure driven flows, the magnitude of the total mechanical energy loss of unit volumetric fluid along the streamwise direction equals to the derivatives of the total mechanical energy in the streamwise direction, that is

$$\frac{\partial H}{\partial s} = -\frac{\partial E}{\partial s}. \qquad (4)$$

For shear driven flows, the derivatives of the total mechanical energy in the transverse direction is the same as Eq.(2). The energy loss of unit volumetric fluid along the streamwise direction equals to the derivatives of the total mechanical energy in the streamwise direction plus the work done to the fluid by external,

$$\frac{\partial H}{\partial s} = -\frac{\partial E}{\partial s} + \frac{\partial W}{\partial s}, \tag{5}$$

where $W$ is the work done to the unit volume fluid by external.

For a given base of parallel flow, the fluid particles may move in an oscillatory pattern in the streamwise direction if they are subjected to a disturbance. With the motion, the fluid particle may gain energy ($\Delta E$) via the disturbance, and simultaneously this particle may have energy loss ($\Delta H$) due to the fluid viscosity along the streamline direction. The analysis in [22, 24] showed that the magnitudes of $\Delta E$ and $\Delta H$ determine the stability of the flow of fluid particles. For parallel flows, the relative magnitude of the energy gained from the disturbance and the energy loss due to viscous friction determines the disturbance amplification or decay. Thus, for a given flow, a stability criterion can be written as follow for a half-period,

$$F = \frac{\Delta E}{\Delta H} = \left(\frac{\partial E}{\partial n}\frac{2\overline{A}}{\pi}\right) \bigg/ \left(\frac{\partial H}{\partial s}\frac{\pi}{\omega_d}u\right) = \frac{2}{\pi^2}K\frac{\overline{A}\omega_d}{u} = \frac{2}{\pi^2}K\frac{v'_m}{u} < Const, \tag{6}$$

and

$$K = \frac{\partial E/\partial n}{\partial H/\partial s}. \tag{7}$$

Here, $F$ is a function of coordinates which expresses the ratio of the energy gained in a half-period by the particle and the energy loss due to viscosity in the half-period. $K$ is a dimensionless field variable (function) and expresses the ratio of transversal energy gradient and the rate of the energy loss along the streamline. $E = \frac{1}{2}\rho V^2$ is the kinetic energy per *unit volumetric fluid*, $s$ is along the streamwise direction and $n$ is along the transverse direction. $H$ is the energy loss per *unit volumetric fluid* along the streamline for finite length. Further, $\rho$ is the fluid density, u is the streamwise velocity of main flow, $\overline{A}$ is the amplitude of the disturbance distance, $\omega_d$ is the frequency of the disturbance, and $v'_m = \overline{A}\omega_d$ is the amplitude of the disturbance of velocity.

Further, $\Delta E = \dfrac{\partial E}{\partial n}\dfrac{2\overline{A}}{\pi}$ and $\Delta H = \dfrac{\partial H}{\partial s}\dfrac{\pi}{\omega_d}u$ are the gradient of total mechanical energy of unit volumetric fluid in the transverse direction and the loss of total mechanical energy of unit volumetric fluid in the streamwise direction, respectively. It can be found from Eq.(6) that the large the value of $F$, the flow is more unstable. There is a constant of $F$ below which the flow remains stable. For given disturbance, the value of K in the flow field determines the flow stability. As stated earlier, the ratio K is a dimensionless function of the flow field, i.e., a function of coordinates (x,y,z). Since K may vary with the flow parameters and the spatial space, the maximum of K in the domain, i.e., $K_{max}$, should bound the stability of the flow for given disturbance. As such, a critical value of $K_{max}$ can be used to expresses the critical condition, and is given as $K_c$.

The energy gradient theory as described for parallel flows in detail in [21,22], can be extended to the curved flow [24], if we change the Cartesian coordinates (x, y) to curvilinear coordinates (s, n), to change the kinetic energy ($\frac{1}{2}mu^2$) to the total mechanical energy ($E = p + \frac{1}{2}\rho u^2$) (the gravitational energy is neglected) in the analysis for the evolution of the disturbed fluid particle, and to make the velocity (*u*) along the streamline direction [24]. Here, p is the hydrodynamic pressure. Thus, after these substitutions, the equation (6) and (7) are also applicable for curved flows. These equations can be derived from the first principle via the same steps as in [22].

In term of Eqs.(6) and (7), the distribution of *K* in the flow field and the property of disturbance may be the perfect means to describe the disturbance amplification or decay in the flow. According to this theory, it can be found that the flow instability first occur at the position of $K_{max}$, for given disturbance, which is construed to be the most "dangerous" position. Thus, for a given disturbance, the occurrence of instability depends on the magnitude of this dimensionless variable *K* and the critical condition is determined by the maximum value of *K* in the flow. For a given flow geometry and fluid properties, when the maximum of *K* in the flow field exceeds a critical value $K_c$, it is expected that instability can occur for a certain initial disturbance [21,22]. Turbulence transition is a local phenomenon in the earlier stage, as found in experiments [4]. For a given flow, K is proportional to the global Reynolds number [21]. A large value of K has the big ability to amplify the disturbance, and vice versa. The energy gradient analysis has suggested that the transition to turbulence is due to the energy gradient and the disturbance amplification [21,22], rather than just the linear eigenvalue instability type as expounded and stated in [26, 27]. Both Trefethen et al. [26] and Grossmann [27] commented that the nature of the onset-of-turbulence mechanism in parallel shear flows must be different from an eigenvalue instability of linear equations of small disturbance. In fact, finite disturbance is needed for the turbulence initiation in the range of finite Re as found in experiments [28]. Dou [21,22] demonstrated that the criterion obtained has a consistent value at the subcritical condition of transition determined by the experimental data for plane Poiseuille flow, pipe Poiseuille flow as well as plane Couette flow (see Table 1). For plane Poiseuille flow, both the two definitions of Reynolds number are given in Table 1 because different definitions are found in literature. In linear stability analysis, $\text{Re} = \rho u_0 l / \mu$ is generally used. Here, $u_0$ is the velocity at the centerline and $l$ is the half width of the channel. Another definition of Reynolds number, $\text{Re} = \rho U L / \mu$, is an analogy to that used in pipe Poiseuille flow. Here, $U$ is the average velocity in the channel and $L$ is the width of the channel.

| Flow type | Re expression | Eigenvalue analysis, $\text{Re}_c$ | Experiments, $\text{Re}_c$ | $K_{max}$ at $\text{Re}_c$ (from experiments), $\equiv K_c$ |
|---|---|---|---|---|
| Pipe Poiseuille | $\text{Re} = \rho U D / \mu$ | Stable for all Re | 2000 | 385 |
| Plane Poiseuille | $\text{Re} = \rho U L / \mu$ | 7696 | 1350 | 389 |
|  | $\text{Re} = \rho u_0 l / \mu$ | 5772 | 1012 | 389 |
| Plane Couette | $\text{Re} = \rho U l / \mu$ | Stable for all Re | 370 | 370 |

Table 1 Comparison of the critical Reynolds number and the energy gradient parameter $K_{max}$ for plane Poiseuille flow and pipe Poiseuille flow as well as for plane Couette flow [21,22]. $U$ is the averaged velocity, $u_0$ the velocity at the mid-plane of the channel, $D$ the diameter of the pipe, $h$ the half-width of the channel for plane Poiseuille flow (L=2l) and plane Couette flow. For Plane Poiseuille flow and pipe Poiseuille flow, the $K_{max}$ occurs at y/l=0.5774, and r/R=0.5774, respectively. For Plane Couette flow, the $K_{max}$ occurs at y/l=1.0.

It can be deduced from Table 1 that the turbulence transition takes place at a consistent critical value of $K_c$ at about 385-389 for both the plane Poiseuille flow and pipe Poiseuille flow, and about 370 for plane Couette flow. This may suggest that the subcritical transition in parallel flows takes place at a consistent value of $K_c \approx 370\text{-}385$. This finding further suggests that the mechanism of flow instability occurring in these basic flows might be the same.

In all the parallel flows, observation can be made that the transverse velocity is v=0 and the pressure is constant in the transverse direction. The variation of total mechanical energy in the transverse direction is only due to the kinetic energy $\frac{1}{2}\rho u^2$ (when the gravitational energy is neglected). Therefore, the gradient of the kinetic energy is the possible source of amplification of disturbance along the transverse direction. In the streamwise direction, the kinetic energy is constant, the energy loss is the pressure drop for pressure driven flows or the input of the external work for shear driven flows, which sustains the velocity profile to keep it constant in the streamwise direction for laminar flows. Due to zero transversal velocity, the diffusion of energy in transverse direction is zero. Therefore, for all the three parallel flows, including pressure driven flow and shear driven flows, the gradient of kinetic energy in transverse direction and the energy loss along the streamline direction are the dominating factors for the flow stability. As such, this can be understood that the mechanism of flow instability in these parallel flows is the same.

It is also noticed that the critical condition for flow instability as determined by linear stability analysis differs largely from the experimental data for all the three different types of flows, as shown in Table 1. Therefore, linear stability analysis is not a good method to analyze the condition for *transition to turbulence*. Using energy gradient theory, it is observed that the balance of the energy amplification in transverse direction and the energy loss in streamwise direction really dominate the flow stability. It is also demonstrated that the viscous flow with an inflectional velocity profile is unstable for both two-dimensional flow and axisymmetric flow [29].

From above discussions, for the plane Poiseuille flow, this said position where $K_{max} > K_c$ should then be the most dangerous location for flow breakdown, which has been confirmed by Nishioka et al's experiment [30]. Nishioka et al's [30] experiments for plane Poiseuille flow showed details of the outline and process of the flow breakdown. The measured instantaneous velocity distributions indicate that the first oscillation of the velocity occurs at y/h=0.50~0.62, as shown by the Fig. 14 in [30]. Nishioka et al. [30] measured the distribution of the instantaneous averaged velocity in a period, and the results indicate that the oscillation of the velocity always occurs in the range of y/h=0.50~0.62 for the disturbance imposed (the base flow keeps laminar).

For the pipe flow, in a recent study, Wedin and Kerswell [31] showed the presence of a "shoulder" in the velocity profile at about r/R=0.6 from their traveling wave solution. They suggested that this position corresponds to where the fast streaks of traveling waves reach the wall. It can be construed that this kind of velocity profile as obtained by simulation is similar to that found in Nishioka et al's experiments for channel flows [30]. The location of the "shoulder" is about the same as that for $K_{max}$ (at y/h=0.5774). According to the present theory, this "shoulder" may then be intricately related to the energy gradient distribution. The solution of traveling waves has been confirmed by experiments recently [32].

In summary, the mechanism for instability described by the function K is that it represents the balance between the two roles of disturbance amplification by the energy gradient in the transverse direction and disturbance damping by the energy loss in the streamwise direction.

## 3. Energy Gradient Theory Applied to Taylor-Couette Flow

We shall assume that the disturbance to the base flow is periodic and wave length is relatively small compared with the scale of the flow geometry. The base flow is assumed to be steady laminar flow. Whether stability criteria are written for the half-period or whole period would be the same since the two half-period are skew-symmetrical in a period. For the wall bounded flows considered here, the boundary conditions is non-slip. For the Taylor-Couette flow, assumptions on the base flow is that expressed by the basic solution (the following Eq.(8-11). The disturbance is assumed as periodic along the streamwise direction of the basic flow (ie., along the circular direction). Under these assumptions, the same expression given previously as Eqs.(6) and (7) can be derived for the circular flow between concentric cylinders [24].

### 3.1 Velocity distribution for Taylor-Couette Flow

The solution of velocity distribution between concentric rotating cylinders can be found in many texts, e.g. [1-3]. Firstly, we define the components of the velocity in the tangential and radial directions as u and v, respectively. Assuming v=0 and $\frac{\partial}{\partial \theta} = 0$, the Navier-Stokes equations in radial and circumferential directions for steady flows reduce to

$$\rho \frac{u^2}{r} = \frac{dp}{dr}, \qquad (8)$$

and

$$\frac{\partial}{\partial r}\left(\frac{\partial u}{\partial r}+\frac{u}{r}\right)=0. \tag{9}$$

Integrating Eq.(9) and using the boundary conditions gives the solution of the velocity field as,

$$u = Ar + \frac{B}{r}, \tag{10}$$

where

$$A = \omega_1 \frac{(\eta^2 - \lambda)}{\eta^2 - 1} \quad \text{and} \quad B = \omega_1 R_1^2 \frac{(1-\lambda)}{1-\eta^2}. \tag{11}$$

In Eq.(11), $\eta = R_1 / R_2$ and $\lambda = \omega_2 / \omega_1$. $R_1$ is the radius of the inner cylinder and $R_2$ is the radius of the outer cylinder. $\omega_1$ and $\omega_2$ are the angular velocities of the inner and outer cylinders, respectively.

### 3.2 Energy gradient in the transverse direction

The gradient of the total mechanical energy gradient in the transverse direction is

$$\frac{\partial E}{\partial r} = \frac{\partial (p + 1/2\rho u^2)}{\partial r} = \rho u \frac{du}{dr} + \rho \frac{u^2}{r}. \tag{12}$$

Introducing Eq.(10) and Eq.(11) into Eq.(12), the energy gradient in the transverse direction therefore is

$$\frac{\partial E}{\partial r} = \rho\left[\left(Ar+\frac{B}{r}\right)\left(A-\frac{B}{r^2}\right)+\frac{1}{r}\left(Ar+\frac{B}{r}\right)^2\right] = 2\rho A\left(Ar+\frac{B}{r}\right). \tag{13}$$

### 3.3 Energy Loss Distribution for Taylor-Couette Flow

The following equation for calculating the radial distribution of rate of energy loss along the streamline for Taylor-Couette flow is obtained as [33],

$$\frac{dH}{ds} \equiv \frac{\tau}{u}\frac{du}{dr} - \frac{\tau}{r}, \tag{14}$$

where $\tau$ is the shear stress. Equation (14) is applicable to flows for one cylinder rotating and the other at rest, and cylinders rotating in opposite directions. For cylinders rotating in the same direction, a different equation must be used [33]

$$\frac{dH}{ds} \equiv \frac{\tau_a}{u_a}\frac{du_a}{dr} - \frac{\tau_a}{r}, \tag{15}$$

where $u_a$ is the velocity in the flow field expressed by $u_a = u - r\omega_2$ assuming that $\omega_1 > \omega_2$ and $\tau_a$ is the shear stress in the velocity field expressed by $u_a$. The details of the derivation for $dH/ds$ can be found in [33] and is not repeated here.

With the velocity gradient obtained from Eq.(10), the shear stress in (14) is therefore,

$$\tau = \mu\left(\frac{\partial u}{\partial r} - \frac{u}{r}\right) = \mu\left[\left(A - \frac{B}{r^2}\right) - \frac{1}{r}\left(Ar + \frac{B}{r}\right)\right] = -\mu\frac{2B}{r^2}, \tag{16}$$

where $\mu$ is the dynamic viscosity. Thus, we have

$$\frac{\tau}{r} = -\mu\frac{2B}{r^3} \tag{17}$$

and

$$\frac{\tau}{u}\frac{du}{dr} = -\mu\frac{2B}{r^2}\left(Ar + \frac{B}{r}\right)^{-1}\left(A - \frac{B}{r^2}\right). \tag{18}$$

Introducing Eqs.(17) and Eq.(18) into Eq.(14), the energy loss is

$$\frac{dH}{ds} \equiv \frac{\tau}{u}\frac{du}{dr} - \frac{\tau}{r} = -\mu\frac{2B}{r^2}\left(Ar + \frac{B}{r}\right)^{-1}\left(A - \frac{B}{r^2}\right) + \mu\frac{2B}{r^3}$$

$$= \mu\frac{2B}{r^2}\left[\frac{1}{r} - \left(Ar + \frac{B}{r}\right)^{-1}\left(A - \frac{B}{r^2}\right)\right] = \mu\frac{4B^2}{r^4}\left(Ar + \frac{B}{r}\right)^{-1}. \tag{19}$$

For cylinders rotating in same direction, using the same procedure as that derived for Eq.(19), the equation can be obtained from Eq.(15) as,

$$\frac{dH}{ds} \equiv \frac{\tau_a}{u_a}\frac{du_a}{dr} - \frac{\tau_a}{r} = \mu \frac{4B_a^2}{r^4}\left(A_a r + \frac{B_a}{r}\right)^{-1}, \tag{20}$$

where

$$A_a = \omega_{1a}\frac{\left(\eta^2 - \lambda_a\right)}{\eta^2 - 1} \quad \text{and} \quad B_a = \omega_{1a} R_1^2 \frac{\left(1 - \lambda_a\right)}{1 - \eta^2}, \tag{21}$$

and $\eta = R_1/R_2$, $\lambda_a = \omega_{2a}/\omega_{1a}$, $\omega_{1a} = \omega_1 - \omega_2$, and $\omega_{2a} = 0$. Here, we have deliberately maintained Eq.(19) and Eq.(20) with similar form for the very purpose that the derivations in subsequent sections below can use essentially the same equation, differing only in the coefficients $A$ and $B$ for Eq.(19) and $A_a$ and $B_a$ for Eq.(20).

### 3.4 Distribution of K

Introducing Eq.(13) and (19 or 20) into Eq.(7), the ratio of the gradients of the total mechanical energy in the transverse direction and the loss of the total mechanical energy in the streamwise direction, K, can be written as,

$$K = \frac{\partial E/\partial r}{\partial H/\partial s} = \frac{\rho u \dfrac{du}{dr} + \rho \dfrac{u^2}{r}}{-\left(\dfrac{\tau}{u}\dfrac{du}{dr} - \dfrac{\tau}{r}\right)} = \frac{1}{\nu}\frac{2A\left(Ar + \dfrac{B}{r}\right)}{\dfrac{4B^{*2}}{r^4}\left(A^*r + \dfrac{B^*}{r}\right)^{-1}}, \tag{22}$$

where $\nu$ is the kinematic viscosity. In this equation, the calculations of A and B are carried out using Eq.(11). The evaluations of $A^*$ and $B^*$ are different for counter rotating and co-rotating cylinders. For cylinders rotating in opposite directions, $A^* = A$ and $B^* = B$ (calculated using Eq.(11)); for cylinders rotating in same direction, $A^* = A_a$ and $B^* = B_a$ (calculated using Eq.(21)).

Introducing Eqs.(10) and (11) or (21) into Eq.(22), then simplifying and rearranging, Eq.(22) becomes,

$$K = \frac{1}{2\nu} \frac{r^4}{R_1^4} \frac{\omega_1}{\omega_1*^2} \frac{(\eta^2-\lambda)(\eta^2-1)}{\omega_1(1-\lambda*)^2} \left[ \frac{\omega_1(\eta^2-\lambda)}{(\eta^2-1)} r - \frac{1}{r}\omega_1 R_1^2 \frac{(1-\lambda)}{(1-\eta^2)} \right] \times$$

$$\left[ \frac{\omega_1*(\eta^2-\lambda*)}{(\eta^2-1)} r - \frac{1}{r}\omega_1*R_1^2 \frac{(1-\lambda*)}{(1-\eta^2)} \right]. \tag{23}$$

The evaluations of $\lambda*$ and $\omega_1*$ are different for counter rotating and co-rotating cylinders. For cylinders rotating in opposite directions, $\lambda* = \lambda$ and $\omega_1* = \omega_1$; For cylinders rotating in same direction, $\lambda* = \lambda_a$ and $\omega_1* = \omega_{1a}$.

Re-arranging, Eq.(23) can be rewritten as

$$K = \frac{1}{2} \frac{\omega_1 R_1^2}{\nu} \frac{\omega_1}{\omega_1*} \frac{r^4}{R_1^4} \frac{(\eta^2-\lambda)}{(1-\lambda*)^2(\eta^2-1)} \times$$

$$\left[ \frac{r}{R_1}(\eta^2-\lambda) - \frac{R_1}{r}(1-\lambda) \right] \left[ \frac{r}{R_1}(\eta^2-\lambda*) - \frac{R_1}{r}(1-\lambda*) \right]. \tag{24}$$

Using a more appropriate form by explicitly showing the Reynolds number, $\text{Re} = \frac{\omega_1 R_1 h}{\nu}$, Eq.(24) can be expressed as

$$K = \frac{1}{2} \text{Re} \frac{R_1}{h} \frac{\omega_1}{\omega_1*} \frac{r^4}{R_1^4} \frac{(\eta^2-\lambda)}{(1-\lambda*)^2(\eta^2-1)} \times$$

$$\left[ \frac{r}{R_1}(\eta^2-\lambda) - \frac{R_1}{r}(1-\lambda) \right] \left[ \frac{r}{R_1}(\eta^2-\lambda*) - \frac{R_1}{r}(1-\lambda*) \right], \tag{25}$$

where $h = R_2 - R_1$ is the gap width between the cylinders.

If the outer cylinder is at rest ($\omega_2 = 0$), and only the inner cylinder is rotating ($\omega_1 \neq 0$), then $\lambda = 0$, $\lambda* = 0$, and $\frac{\omega_1}{\omega_1*} = 1$. Further simplifying Eq.(25), we obtain

$$K = \frac{1}{2} \text{Re} \frac{R_1^2}{h^2} \frac{R_1}{R_1+R_2} \frac{r^2}{R_1^2} \left[ 1 - \frac{r^2}{R_2^2} \right]^2. \tag{26}$$

Next, by letting $r = R_2 - y$, Eq.(26) is rewritten as

$$K = \frac{1}{2}\text{Re}\frac{R_1^2}{h^2}\frac{R_1}{R_1+R_2}\frac{(R_2-y)^2}{R_2^2}\frac{(2R_2-y)^2}{R_2^2}\frac{y^2}{R_1^2}$$

$$= \text{Re}\frac{y^2}{h^2}\frac{R_1}{2(R_1+R_2)}\frac{(R_2-y)^2}{R_2^2}\frac{(2R_2-y)^2}{R_2^2}$$

$$= \text{Re}\frac{y^2}{h^2}\frac{R_1}{2(R_1+R_2)}\left(1-\frac{y}{h}\frac{h}{R_2}\right)^2\left(2-\frac{y}{h}\frac{h}{R_2}\right)^2. \qquad (27)$$

This equation easily relates to plane Couette flow. Plane Couette flow can have two configurations: two plates move in opposite directions and one plate moves while the other is at rest. Taylor-Couette flow with $\omega_2 = 0$ and $\omega_1 \neq 0$ corresponds to plane Couette flow for the latter case. From Eq.(27), it can be seen that K is proportional to Re in any location in the field. K is an eighth order function of distance from the outer cylinder across the channel, which is related to the value of relative channel width $h/R_2$. The distribution of K along the channel width between cylinders calculated using Eq. (27) is depicted in Fig.2 for various values of h/R$_2$ with the inner cylinder rotating while the outer cylinder is kept at rest ($\omega_2 = 0$). For a given Re and h/R$_2$, it is found that K increases with increasing y/h and the maximum of K is obtained at $y/h = 1$ for low values of h/R$_2$ (h/R$_2$<0.43). That is, it reaches its maximum at the surface of the inner cylinder. For higher value of h/R$_2$ (h/R$_2$>0.43), the location of K$_{max}$ moves to within the flow located between y/h=0 and y/h=1. The cases studied in the literature are usually for low gap width. We shall focus our discussion for the case of h/R$_2$<0.43 in this study. The maximum of K for h/R$_2$<0.43 can be expressed as

$$K_{max} = \text{Re}\frac{R_1}{2(R_1+R_2)}\left(1-\frac{h}{R_2}\right)^2\left(2-\frac{h}{R_2}\right)^2. \qquad (28)$$

It is found from Eq. (28) that K$_{max}$ depends on Reynolds number and the geometry. As we will see below, the critical stability condition will be determined by Eq.(28). When the cylinder radii tend to infinity, we have in Eq.(27)

$$\frac{R_1}{2(R_1+R_2)} \to \frac{1}{4}, \left(1-\frac{y}{h}\frac{h}{R_2}\right)^2 \to 1, \text{ and } \left(2-\frac{y}{h}\frac{h}{R_2}\right)^2 \to 4. \tag{29}$$

Then, Eq.(27) reduces to

$$K = \text{Re}\frac{y^2}{h^2}. \tag{30}$$

This equation at the limit of infinite radii of cylinders is the same as that for plane Couette flow [23]. The corresponding maximum of K at y=h is

$$K_{max} = \text{Re} = \frac{\omega_1 R_1 h}{\nu}. \tag{31}$$

As discussed in [22, 24], the development of the disturbance in the flow is subjected to the mean flow condition and the boundary and initial conditions. The mean flow is characterized by energy gradient function K. Therefore, the flow stability depends on the distribution of K in the flow field and the initial disturbance provided to the flow. In the flow regime with high value of K, the flow is more unstable than that in the flow regime with low value of K. The first sign of instability should be associated with the maximum of K ($K_{max}$) in the flow field for a given disturbance. In other words, the position of maximum of K is the most dangerous position. For given flow disturbance, there is a critical value of $K_{max}$ over which the flow becomes unstable. It is not trivial to directly predict this critical value $K_c$ by theory as in parallel flows [21] since it is obviously a strongly nonlinear process and the usual tool for perturbation analysis is not applicable. Nevertheless, it can be observed in experiments as those done for parallel flows. The value of $K_{max}$ when flow instability occurs can be taken as a criterion for instability, and this value is recorded as $K_c$; if $K_{max} > K_c$, the flow will become unstable.

Thus, the study of distribution of K in the flow field can help to locate the region where the flow is inclined to be unstable. In Fig.2, K increases with increasing y/h for given $h/R_2$ (at low value of $h/R_2$), and its maximum occurs at the inner cylinder. Thus, the flow at the outer cylinder is most stable and the flow at the inner cylinder is most unstable. Therefore, a small disturbance can be amplified at the inner cylinder if the value of K reaches its critical value for the given geometry. In other words, the inner cylinder is a possible location for first occurrence of instability, as generally observed in the experiments [5,16].

In Fig.2, the line for $h/R_2=0$ corresponds to plane Couette flow wherein, one plate moves while the other is at rest, which is a parabola (i.e., Eq.(30)) [23]. It can be found that there is little difference in the distribution of K for $h/R_2=0.01$ and $h/R_2=0$. In terms of that view, one may expect that the critical conditions of instability for these two values of $h/R_2$ close to one another. When $h/R_2$ increases, $K_{max}$ decreases. This does not, however, imply that the flow becomes more stable as $h/R_2$ increases. This is because the critical value of $K_{max}$ varies with the variation of $h/R_2$. It will be shown by experimental data in subsequent sections that $K_c$ decreases with the increasing $h/R_2$.

## 4. Comparison with Experiments at Critical Condition

Taylor [5] used a graph of $\omega_1/\nu$ versus $\omega_2/\nu$ to present the results of the critical condition for the primary instability. In order to use the same chart as Taylor for ease of reference, the comparison of theory with experiments is also plotted in this way.

Rewriting Eq.(24), we have

$$\left(\eta^2-\lambda\right)=\frac{K}{\frac{1}{2}\frac{\omega_1 R_1^2}{\nu}\frac{\omega_1}{\omega_1^*}\frac{r^4}{R_1^4}}\frac{\left(1-\lambda^*\right)^2\left(\eta^2-1\right)}{\left[\frac{r}{R_1}\left(\eta^2-\lambda\right)-\frac{R_1}{r}(1-\lambda)\right]\left[\frac{r}{R_1}\left(\eta^2-\lambda^*\right)-\frac{R_1}{r}(1-\lambda^*)\right]}. \quad (32)$$

Rearranging Eq.(32), the following equation (33) is obtained,

$$\frac{\omega_1}{\nu}=\frac{\omega_2}{\nu}\frac{R_2^2}{R_1^2}+\frac{2K}{R_2^2}\frac{\omega_1^*}{\omega_1}\frac{R_2^4}{r^4}\frac{\left(1-\lambda^*\right)^2\left(\eta^2-1\right)}{\left[\frac{r}{R_1}\left(\eta^2-\lambda\right)-\frac{R_1}{r}(1-\lambda)\right]\left[\frac{r}{R_1}\left(\eta^2-\lambda^*\right)-\frac{R_1}{r}(1-\lambda^*)\right]}. \quad (33)$$

Thus, the critical condition for a given geometry is given by $K_c$. That is

$$\left(\frac{\omega_1}{\nu}\right)_c=\frac{\omega_2}{\nu}\frac{R_2^2}{R_1^2}+\frac{2K_c}{R_2^2}\frac{\omega_1^*}{\omega_1}\frac{R_2^4}{r^4}\frac{\left(1-\lambda^*\right)^2\left(\eta^2-1\right)}{\left[\frac{r}{R_1}\left(\eta^2-\lambda\right)-\frac{R_1}{r}(1-\lambda)\right]\left[\frac{r}{R_1}\left(\eta^2-\lambda^*\right)-\frac{R_1}{r}(1-\lambda^*)\right]}.$$

(34)

In Equation (34), $K_c$ is the critical value of $K_{max}$ at the primary instability condition, which can be determined from experiments. For a given flow geometry, $K_c$ is treated as constant for the initiation of instability as described before. After the value of $K_c$ is determined, the value of $(\omega_1/\nu)_c$ can be solved by an iteration process for an initial value of $\lambda$ and a given value of $\omega_2/\nu$. The calculated results with the theory are compared with available experimental data in literature [5][11][13][16] concerning the primary instability condition of Taylor-Couette flow. Figures 3 to 4 show the comparison of theory with Taylor's experiments [5] for two parametric conditions, while Fig. 5, Fig.6 and Fig.7 show the comparisons of theory with Coles' experiments [13], Snyder's experiments [11], and Andereck et al's experiments [16], respectively. In these figures, the critical value of the energy gradient parameter K ($K_c$) is determined by the experimental data at $\omega_2 = 0$ and $\omega_1 \neq 0$ (the outer cylinder is fixed, the inner cylinder is rotating). Using the determined value of $K_c$ for a given set of geometrical parameters, the critical value of $\omega_1/\nu$ versus $\omega_2/\nu$ is calculated for a range of $\omega_2/\nu$ as in the experiments using Eq.(34). In [12], the constant in the analytical equation obtained was fixed using the result of linear stability analysis.

It can be seen from Figs.3-7 then that when the cylinders rotate in the same direction, the theory obtains very good concurrence with all the experimental data. When cylinders rotate in opposite directions, the theory obtains good agreement with the experimental data for small relative gap width ($h/R_1$). For larger relative gap width, the theory has some deviations from the experimental data with increasing negative rotation speed of the outer cylinder. The reason can be explained as follows. When the gap is large and the cylinders are rotating in opposite directions, the flow in the gap is more distorted compared to plane Couette flow (linear velocity distribution). This distortion of velocity profile has an effect on the distributions of flow energy loss and energy gradient. The magnitude of flow energy loss dH/ds can be calculated by Eq.(19 or 20) and the energy gradient can be calculated by Eq.(13). On the other hand, if the rotating speed of the outer cylinder is high, the flow layer near the outer cylinder may earlier transit directly to turbulence if the disturbance is sufficiently large [6][13], which has not been the focus of researches before. This will obviously alter the velocity profile of the flow and influence the distribution of the energy gradient function K and the maximum of K (more discussion will be given in the paragraph below when Fig.8 is introduced and discussed). For example, in Andereck et al's experiments [16], when $\omega_2/\nu$ is -100 and the inner cylinder is at rest, the Reynolds number based on the rotation speed of the outer cylinder $Re_2$ ( $Re_2 = R_2 h \omega_2/\nu$ ) reaches 416. At this value of 416, plane Couette flow has already become turbulent ($Re_c$=325--370). For counter-rotating cylinders with curved streamlines, the transition must occur earlier than that in plane Couette flow because of the influence of the radial pressure gradient which increases the radial (transverse) energy gradient near the outer cylinder. The same type of deviation in prediction is also observed in the comparison of Taylor's mathematical theory with his experiments when cylinders rotate in opposite directions at large negative rotating speed of outer cylinder; in particular, if the relative gap is large [5]. Therefore, when cylinders rotate in opposite directions, further study is needed to study the occurrence of turbulence as induced by shear flow near the outer cylinder (caused by convective inertia). This is compared with the Taylor vortex pattern as induced by the centrifugal force near the inner cylinder when only the inner cylinder is rotating.

It is found that the first term on the right hand side of Eq.(34) is that for Rayleigh's inviscid criterion, and the second term on the right hand side of Eq.(34) is due to the effect of viscous friction. If $K_c$ is zero, Eq.(34) degenerates to the Rayleigh's equation. In Figs.3-7, Rayleigh's inviscid criterion ($(\omega_1)_c = (R_2/R_1)^2 \omega_2$) is also included for comparison. In Taylor's calculations and experimental results [5], it was shown that viscosity has only stabilizing role to the flow between the concentric cylinders. For cylinders rotating in the same direction, our theory shows very good agreement with the experiments and demonstrates that viscosity has a stabilizing effect on the flow, as compared to the inviscid case. For inviscid flow at $\omega_2/\nu$=0, $(\omega_1/\nu)_c$=0. For the viscous flow, the viscosity plays a stabilizing role and hence gives rise to a certain threshold quantity of $(\omega_1/\nu)_c$ not equal zero, below which the flow is stable. The mechanism of the stability role of viscosity is the same as for the parallel flows [21-24]. Viscosity leads to energy loss along the flow path which leads to the disturbance damping when the disturbance propagates between the fluid layers. Thus, viscosity increases the stability of shear flows and enhances the value of $(\omega_1/\nu)_c$.

In Fig.8, we show the distribution of K along the channel width at the critical condition of $K_c$=77 as shown in Fig.4, which is taken from Taylor's experiments for $R_1$=3.55cm and $R_2$=4.035 cm. The value of $K_c$ is only dependent on the ratio of radius, and is independent of the rotating speed of cylinders. It can be seen in Fig.8a that K increases monotonically from the outer cylinder to the inner cylinder, when the inner cylinder is rotating while the outer cylinder is at rest. The maximum of K occurs at the inner cylinder, so the stability of the flow is dominated by the $K_{max}$ at the inner cylinder. In Fig.8b, it can be seen that K increases monotonically from the outer cylinder to the inner cylinder, when the two cylinders are rotating in same direction and $\omega_1/\nu$ is larger than $\omega_2/\nu$. The maximum of K also occurs at the inner cylinder, so the stability of the flow is dominated by the $K_{max}$ at the inner cylinder too. In these two pictures, the base flow in the gap is laminar flow. Taylor vortex cell pattern are found in these cases as shown in experiments [5,16]. When the two cylinders rotate in opposite directions, the distribution of K generates two maxima respectively at the inner cylinder and the outer cylinder. In Fig.8c, it can be seen that the maximum at the outer cylinder is not high since the speed of the outer cylinder is small. In this case the base flow in the gap may be still laminar, and the stability of the flow is still completely dominated by the $K_{max}$ at the inner cylinder. If the speed of the outer cylinder becomes high and exceeds certain critical value, the flow near the outer cylinder may become turbulence provided that the disturbance is sufficiently large [6][13]. As shown in Fig.8d, the value of K at the outer cylinder (K=367) is about or higher than the critical value for plane Couette flow to transit to turbulence ($K_c$=325-370), the flow layer near the outer cylinder may already be turbulent. Thus, although the base flow is laminar near the inner cylinder but the flow may have transited to turbulence near the outer cylinder, at this flow state which is the critical condition of primary instability deemed dominated by the rotation of inner cylinder. Therefore, as this critical condition is exceeded with increasing $\omega_1/\nu$, but when $\omega_2/\nu$ is large and negative, the usual Taylor vortex cell pattern may not materialize; instead spiral turbulence is generated [13,16]. This is because the generation of turbulence near the outer cylinder has altered the velocity distribution from its original laminar behaviour. The circulation of fluid particle between the two cylinder surfaces (alternating between the laminar region and the turbulent region) forms an intermittent and spiral turbulence pattern. This may provide a plausible explanation for the observation of spiral turbulence pattern as found in experiments [13][16]. As reproduced in Fig.9, Andereck et al [16] plotted regimes of the flows in terms of Ro and Ri as coordinates (shown as Fig.1 in their paper). Here Ro and Ri are the Reynolds number based on the rotating speed of outer and inner cylinders, respectively. The behaviour of the flow may be better explained using the distribution of K along the gap width, as discussed above.

In Figs.10 and 11, we show the isolines of the $K_{max}$ along the side of inner cylinder in the plane of $\omega_1/\nu$ versus $\omega_2/\nu$ which occurs on the surface of the inner cylinder. Because the energy gradient dominates the flow behaviour and controls the mechanism of the flow instability and transition, the classification of regimes may be better understood using the isolines shown in Figs.10-11. By comparing Figs.10-11 and Fig.9, the regimes of the flow from experiments may appear to be aptly characterized by the isolines of $K_{max}$ along the inner cylinder in $\omega_1/\nu$ and $\omega_2/\nu$ plane. It should be noticed that the isolines of $K_{max}$ for $K_{max}<K_c$ are exact, while the isolines for $K_{max}>K_c$ are approximate because the velocity distribution in the gap can not be accurately expressed by Eq.(10) anymore due to the formation of Taylor vortices or spiral vortices/ spiral turbulence. It should be made clear that $K_{max}$ is the maximum of the magnitude of K in the flow domain at a given $\omega_1/\nu$ and $\omega_2/\nu$ condition and geometry, and $K_c$ is critical value of $K_{max}$ at the primary instability for a given geometry.

It would be (most) interesting to obtain a unified description for rotating flows and parallel flows vis-a-vis the mechanism of instability. As introduced before, although the critical Reynolds number differs greatly in magnitude for plane Couette flow, plane Poiseuille flow and pipe Poiseuille flow, the critical value of the $K_{max}$ is about the same for all the three mentioned kinds of flows (325-389). Plane Couette flow is the limiting case of Taylor-Couette flow when the curvature of walls is zero. The limiting value of critical condition of Taylor-Couette flow should be the same as that for plane Couette flows. Lundbladh and Johansson's direct numerical simulation produced a critical condition of Rec=375 for plane Couette flow [34]. Another three (independent) research groups also obtained Rec=$370 \pm 10$ in experiments via flow visualization technique during the period 1992-1995 [35-37]. Some subsequent experiments showed a lower critical Reynolds number of 325 [38-39]. In order to include all possible results, the data can be classified as in the range of 325-370 for plane Couette flow. Our derivation has shown that $K_{max}$=Re for plane Couette flow as indicated by Eq.(30). Using these data for Rec, the critical value of $K_{max}$ for plane Couette flow is taken to be $K_c$ =325-370, below which no turbulence occurs regardless of the disturbance.

| Authors | $R_1$ (cm) | $R_2$ (cm) | h (cm) | h/$R_1$ | $(\omega_1/\nu)_c$ (cm$^{-2}$) | Re$_c$ | K$_c$ |
|---|---|---|---|---|---|---|---|
| Taylor (1923) | 3.80 | 4.035 | 0.235 | 0.06184 | 189.2 | 169 | 139 |
| | 3.55 | 4.035 | 0.485 | 0.1366 | 70.7 | 120 | 77 |
| | 3.00 | 4.035 | 1.035 | 0.345 | 30.5 | 95 | 33 |
| Coles (1965) | 10.155 | 11.52 | 1.365 | 0.1343 | 8.4 | 116 | 75 |
| Snyder (1968) | 6.023 | 6.281 | 0.258 | 0.0428 | 139.9 | 217 | 188 |
| | 5.032 | 6.281 | 1.249 | 0.248 | 15 | 94 | 44 |
| Gollub & Swinney (1975) | 2.224 | 2.540 | 0.316 | 0.142 | 182. | 128 | 80 |
| Andereck et al (1986) | 5.25 | 5.946 | 0.696 | 0.1326 | 33. | 120 | 78 |
| Hinko (2003) | 29.54 | 29.84 | 0.30 | 0.01 | 39.5 | 350 | 338 |
| Prigent & Dauchot (2004) | 4.909 | 4.995 | 0.0863 | 0.01752 | 758 | 320 | 301 |

Table 2 Collected data for the detailed geometrical parameters for the experiments and the critical condition determined for the case of the outer cylinder at rest ($\omega_2 = 0$) and the inner cylinder rotating ($\omega_1 \neq 0$).

In Table 2, experimental data are collated for the critical condition of the primary instability in the Taylor-Couette flows. A most interesting result for small gap flow was obtained by Hinko [18] recently. This result is useful to clarify how the Taylor-Coutte flow is related to plane Couette flow. Hinko obtained $Re_c$=350 for the flow in small gap of concentric rotating cylinders with $h/R_1$=0.01. Under this critical condition, the Taylor number is T=$350^2$✕0.01=1225. This value is quite different from the generally acceptable theoretical value of 1708. For this experiment, $K_c$=338 is obtained using Eq.(28). This value approaches the critical value for plane Couette flow of 325-370. All the experimental data for the primary instability in Taylor-Couette flows are depicted in Fig.12 by plotting $K_c$ versus the relative gap width $h/R_1$. The critical value $K_c$ of $K_{max}$ for plane Couette flow, plane Poiseuille flow and pipe Poiseuille flow are also included with $h/R_1$=0. For all the wall-bounded parallel flows, $K_c$=325—389, which are calculated from the experimental data [21, 22, 23]. It is noted from Fig.12 that $K_c$ decreases with increasing $h/R_1$, which depends on Re and $h/R_1$ as also shown by Eq.(28). When $h/R_1$ tends to zero, the value of $K_c$ tends to the value of plane Couette flow.

It may be observed from Fig.12 that there seems a correlation/relationship for the K curve for all types of wall-bounded flows (including Taylor-Couette flows, plane Couette flow, plane Poiseuille flow, and pipe Poiseuille flow). However, it should be mentioned that the critical condition from experiments for Taylor-Couette flow is the primary instability for cell pattern formation which corresponds to the *infinitesimal disturbance* (which is consistent with the prediction by linear stability analysis [5]), while the critical conditions for wall bounded parallel flows were those to sustain the turbulence (spot) below which no turbulence can be generated which corresponds to *finite amplitude disturbance*. This difference pertaining to the initial disturbance needs to be further investigated in future.

Taylor number has been used to describe the stability of Taylor-Couette flow as discussed in the introduction [1-5]. The Taylor number for the case of the outer cylinder fixed ($\omega_2 = 0$) is, $T = \text{Re}^2 (h/R_0)$, with $\text{Re} \equiv \dfrac{\omega_1 R_1 h}{\nu}$. The critical value for instability is Tc=1708 from linear stability calculation [1-2]. When $h/R_0$ tends to zero, the flow reduces to plane Couette flow. In terms of the Taylor number, when $h/R_0$ tends to zero ($R_0$ tends to infinite), T=0 and definitely T<Tc; this means that the flow is always stable. In other words, by stating Tc=1708, the critical Re is infinite if $h/R_0$ tends to zero. This contradicts the experimental results of plane Couette flow. Obviously, if the Taylor-Couette flow is related to plane Couette flow, then the Taylor number may not be sufficient or appropriate to describe the transition. It is only applicable for concentric rotating cylinders with the magnitude of $h/R_0$ not very large or very small.

Taylor [5] used mathematical theory and linear stability analysis and showed that linear stability theory agrees well with experiments. However, the linear stability theory presupposes an infinitesmall disturbance as introduced for the (rotating) Taylor-Couette flow just as for the emperiments. For the parallel flows, the stability conditions were obtained at finite amplitude of disturbance as for the associated experiments. A unified picture linking the stability criterion/criteria established between the Taylor-Couette flow and parallel flows may be incongruous. On the other hand, as shown in this paper, the present theory is valid for all of these concerned flows. Therefore, it is postulated that the energy gradient theory is at the very least a more feasible universal theory for flow instability and turbulent transition, and which is valid for both pressure and shear driven flows in both parallel flow and rotating flow configurations.

## 5. Conclusion

In this paper, the energy gradient theory is applied to Taylor-Couette flow between concentric rotating cylinders. The derivation for the energy gradient function K is given for Taylor-Couette flow, which is also related to plane Couette flow. The limit of infinite cylinder radii of Taylor-Couette flow corresponds to plane Couette flows. The theoretical results for the critical condition found have very good concurrence with the experiments in the literature. The conclusions drawn are:

(1) The energy gradient method as a semi-empirical theory is valid for rotating flows. The critical value of $K_{max}$ for primary instability in Taylor-Couette flows is a constant for a given geometry as confirmed by the experimental data. Therefore, this may suggest that the energy gradient function K is a very reasonable parameter to describe the instability in Taylor-Couette flow.

(2) The isoline chart on the plane of $\omega_1/\nu$ versus $\omega_2/\nu$ may provide a basic physical explanation of the regimes of flow patterns found in the experiments of Andereck et al.[16]

(3) All wall-bounded shear flows share the same mechanism for the instability initiation based on the relative dominance between the gradient of total mechanical energy and the loss of total mechanical energy in the flow. The limit of Taylor-Couette flow at very small gap width becomes that of plane Couette flow.

(4) The K function is useful for relating the plane Couette flow to the Taylor-Couette flow. It has a clear physical concept and meaning. On the other hand, Taylor number is not valid or appropriate in the limiting case of Taylor-Couette flow at very small gap width when the radii of cylinders tend towards infinity.

(5) The energy gradient theory can function as a plausible universal theory for flow instability and turbulent transition and which is valid for both pressure and shear driven flows in both parallel flow and rotating flow configurations. It is shown in the present study that there are some similarities between Taylor-Couette flow and planar Couette flow.

**Acknowledgements**
The authors thank Dr. HM Tsai for helpful discussions.

**References**

1. S. Chandrasekhar, Hydrodynamics and Hydromagnetic Stability, Dover, New York, 1961, 272-381.
2. P. G. Drazin and W. H. Reid, Hydrodynamic stability, Cambridge University Press, 2$^{nd}$ Ed., Cambridge, England, 2004, 69-123.
3. P. Chossat and G. Iooss, The Couette-Taylor Problem, Springer-Verlag, 1994.
4. H. Schlichting, Boundary Layer Theory, (Springer, 7th Ed., Berlin, 1979), 83-111; 449-554.
5. G. I. Taylor, Stability of a viscous liquid contained between two rotating cylinders, Philosophical Transactions of the Royal Society of London. Series A, 223 (1923), 289-343.
6. Donnelly, R.J., Taylor-Couette flow: the early days, Physics Today, No.11, 44 (1991), 32-39.
7. R. Tagg, The Couette-Taylor problem, Nonlinear Science Today, 4 (1994) 2-25.
8. M. Brenner, H. Stone, Modern classical physics through the work of G. I. Taylor, Physics Today, No.5, 2000, 30-35.
9. L. Rayleigh, On the dynamics of revolving fluids, Proceedings of the Royal Society of London. Series A, Vol. 93, No. 648. (Mar. 1, 1917), 148-154
10. T. von Karman,1934, Some aspects of the turbulence problem. Proc. 4$^{th}$ Inter. Congr. For applied Mech., Cambridge, England, 54-91. Also Collected works (1956), Vol.3, Butterworths Scientific Publications, London, 120-155.
11. H. A. Snyder, Stability of rotating Couette flow. II. Comparison with numerical results, Phys. Fluids, 11 (1968) 1599-1605.
12. A. Esser and S.Grossmann, Analytic expression for Taylor–Couette stability boundary, Phys. Fluids, 8(7), 1996, 1814-1819.
13. D. Coles, Transition in circular Couette flow, J. Fluid Mech., 21(1965), 385–425.
14. E. R. Krueger, A. Gross & R.C. DiPrima, On the relative importance of Taylor-vortex and non-axisymmetric modes in flow between rotating cylinders, J. Fluid Mech. 24 (1966), 521–538.
15. J.P. Gollub, and H.L.Swinney, Onset of turbulence in a rotating fluid, Phys. Rev. Lett., 35, 1975, 927-930.
16. C.D. Andereck, S.S. Liu, and H.L. Swinney, Flow regimes in a circular Couette system with independently rotating cylinders, J. Fluid Mech., 164 (1986), pp. 155-183.
17. W.F. Langford, R. Tagg, E.J. Kostelich, H.L. Swinney & M. Golubitsky, Primary

instabilities and bicriticality in flow between counter-rotating cylinders, Phys. Fluids, 31 (1998), 776–785.
18. K. A. Hinko, Transitions in the Small Gap Limit of Taylor-Couette Flow, The Ohio State University Physics Summer Institute, REU Summer 2003; Advisor: Dr. C. D. Andereck, Department of Physics, The Ohio State University, 2003.
19. H. Faisst & B. Eckhardt, Transition from the Couette-Taylor system to the plane Couette system, Phys. Rev. E 61 (2000), 7227–7230.
20. A. Prigent, O. Dauchot, "Barber pole turbulence" in large aspect ratio Taylor-Couette flow, arXiv:cond-mat/0009241 v1, 15 Sep 2000.
21. H-S Dou, Mechanism of flow instability and transition to turbulence, International Journal of Non-Linear Mechanics, Vol.41, May 2006, 512-517.
  http://arxiv.org/abs/nlin.CD/0501049,
22. H.-S. Dou, Physics of flow instability and turbulent transition in shear flows, Technical Report, National University of Singapore, 2006.
  http://www.arxiv.org/abs/physics/0607004 , Also as part of the invited Lecturers: H.-S. Dou, Secret of Tornado, International Workshop on Geophysical Fluid Dynamics and Scalar Transport in the Tropics, NUS, Singapore, 13 Nov. - 8 Dec., 2006.
23. H-S. Dou, B.C.Khoo, K.S. Yeo, Turbulent transition in plane Couette flows, Proc. of the Fifth International Conference on Fluid Mechanics, Ed. by F. Zhuang and J. Li, Tsinghua University Press & Springer-Verlag, 2007, pp.77.
   http://arxiv.org/abs/nlin.CD/0501048
24. H.S. Dou, Three important theorems for flow stability, Proc. of the Fifth International Conference on Fluid Mechanics, Ed. by F. Zhuang and J. Li, Tsinghua University Press & Springer-Verlag, 2007, pp.57-60.
    http://www.arxiv.org/abs/physics/0610082
25. H.-S. Dou and N. Phan-Thien, Viscoelastic flows around a confined cylinder, Part one: Instability and velocity inflection, Chemical Engineering Science, Vol.62, No.15, 2007, 3909-3929.
26. L.N.Trefethen, A.E. Trefethen, S.C.Reddy, T.A.Driscoll, Hydrodynamic stability without eigenvalues, Science, 261 (1993), 578-584.
27. S. Grossmann, The onset of shear flow turbulence. Reviews of Modern Physics, 72 (2000), 603-618.
28. A.G.Darbyshire and T.Mullin, Transition to turbulence in constant-mass-flux pipe flow, J. Fluid Mech, 289 (1995), 83-114.
29. H-S Dou, Viscous instability of inflectional velocity profile, Recent Advances in Fluid Mechanics, Proc. of the 4th Inter. Conf. on Fluid Mech., July 20~23, 2004, Dalian, China; Tsinghua University Press & Springer-Verlag, 2004, 76-79.
     http://arxiv.org/abs/physics/0502091
30. M. Nishioka, S Iida, and Y.Ichikawa, An experimental investigation of the stability of plane Poiseuille flow, J. Fluid Mech., 72 (1975), 731-751.
31. H. Wedin, and R.R. Kerswell, Exact coherent structures in pipe flow: travelling wave solutions, J. Fluid Mech. 508 (2004), 333-371.
32. B. Hof, C.W. H. van Doorne, J.Westerweel, F.T. M. Nieuwstadt, H.Faisst, B.Eckhardt, H.Wedin, R.R. Kerswell, F.Waleffe, Experimental observation of nonlinear traveling waves in turbulent pipe flow, Science, 305 (2004), No. 5690, 10 Sep. 2004, 1594-1598.
33. H.-S. Dou, B.C. Khoo, and K.S.Yeo, Energy Loss Distribution in the Plane Couette Flow and the Taylor-Couette Flow between Concentric Rotating Cylinders, International Journal of Thermal Science, Vol.46, 2007, 262-275.
34. A. Lundbladh and A. Johansson, Direct simulation of turbulent spots in plane Couette flow, J. Fluid Mech. 229 (1991), 499--516.
35. N. Tillmark and P. H. Alfredsson, Experiments on transition in plane Couette flow, J. Fluid

Mech. 235, 89 –102, 1992.
36. F. Daviaud, J. Hegseth, and P. Berge′, Subcritical transition to turbulence in plane Couette flow, Phys. Rev. Lett. 69 (1992), 2511-2514.
37. S. Malerud, K. J. Malfy, and W.I. Goldburg, Measurements of turbulent velocity fluctuations in a planar Couette cell, Phys. Fluids, 7 (1995) 1949--1955.
38. O. Dauchot and F. Daviaud, Finite-amplitude perturbation and spots growth mechanism in plane Couette flow, Phys. Fluids, 7 (1995), 335-343.
39. S. Bottin, O. Dauchot, and F. Daviaud, P. Manneville, Experimental evidence of streamwise vortices as finite amplitude solutions in transitional plane Couette flow. Physics of Fluids, 10(1998) 2597-2607

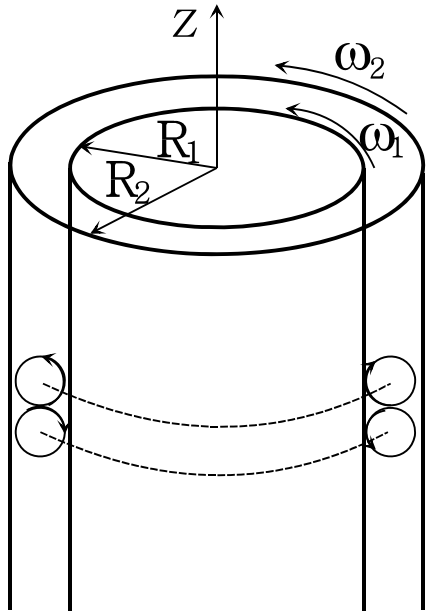

Fig.1 Taylor-Couette flow between concentric rotating cylinders

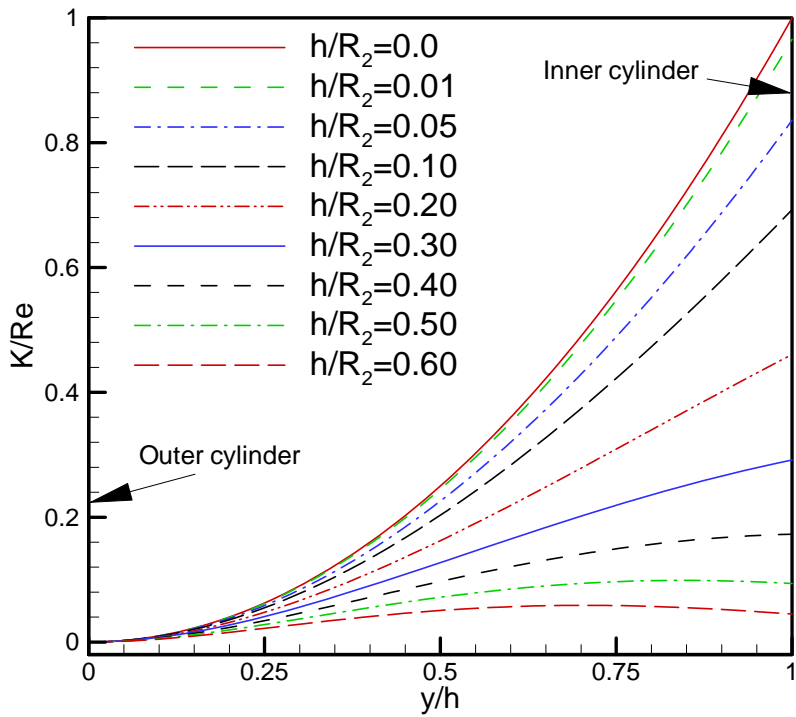

Fig.2 K/Re versus the channel width between the cylinders at various $h/R_2$ for $\omega_2 = 0$ and $\omega_1 \neq 0$ (the outer cylinder is fixed and the inner cylinder is rotating).

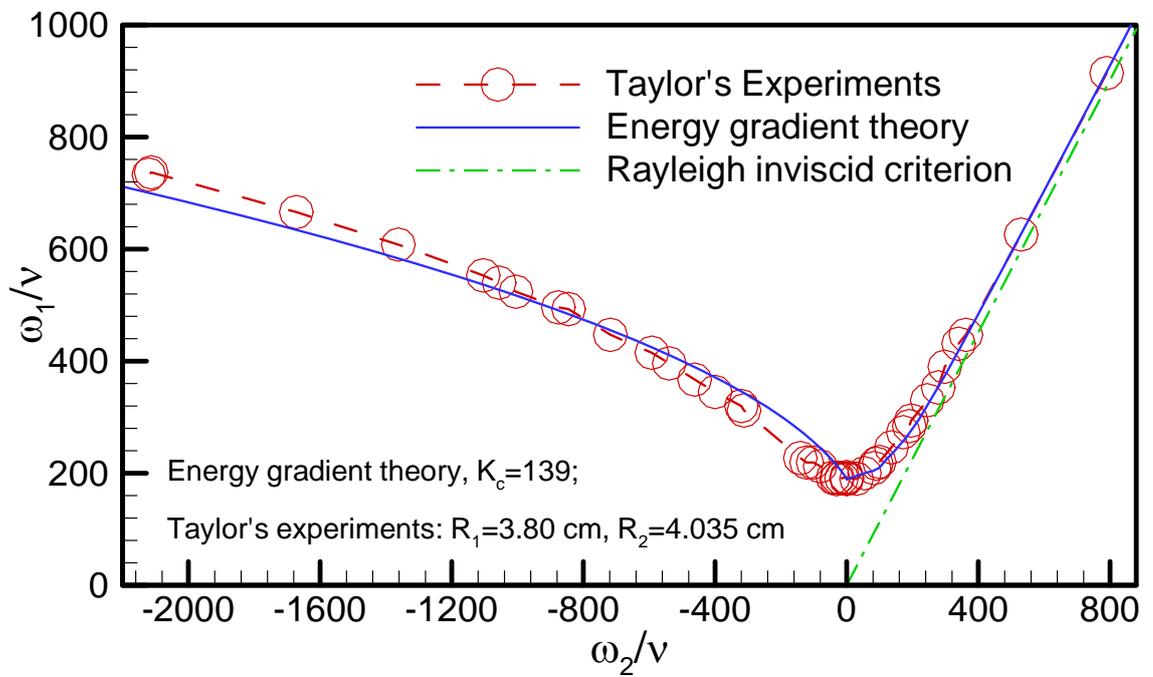

Fig.3 Comparison of the theory with the experimental data for the instability condition of Taylor-Couette flow (Taylor (1923)'s experiments, R1=3.80 cm, R2=4.035 cm). The relative gap width is $h/R_1$=0.06184.

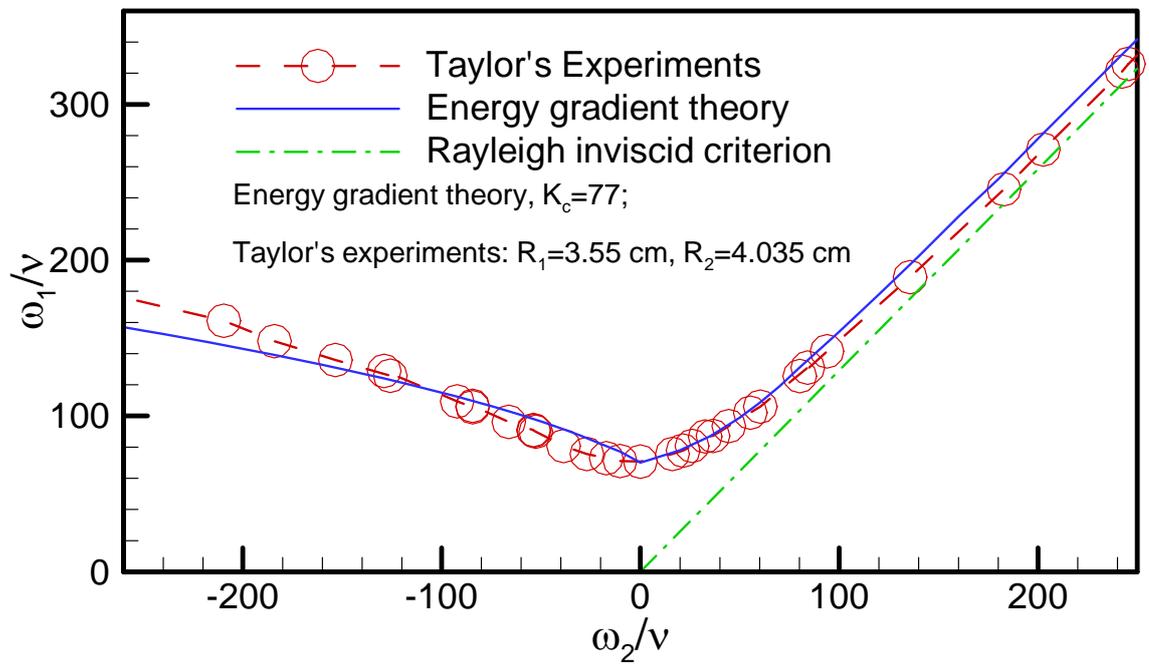

Fig.4 Comparison of the theory with the experimental data for the instability condition of Taylor-Couette flow (Taylor (1923)'s experiments, R1=3.55cm, R2=4.035 cm). The relative gap width is $h/R_1$=0.1366.

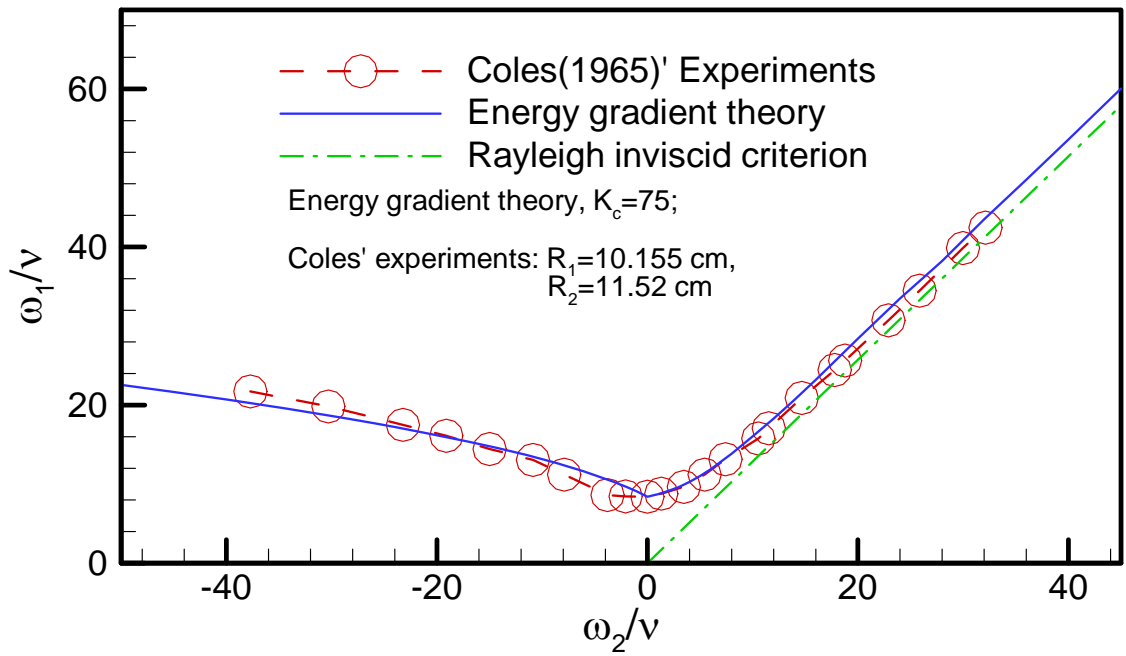

Fig.5 Comparison of the theory with the experimental data for the instability condition of Taylor-Couette flow (Coles (1965)' experiments, R1=10.155 cm, R2=11.52 cm). The data are taken from Fig.2c in the paper [13]. The relative gap width is $h/R_1=0.1343$

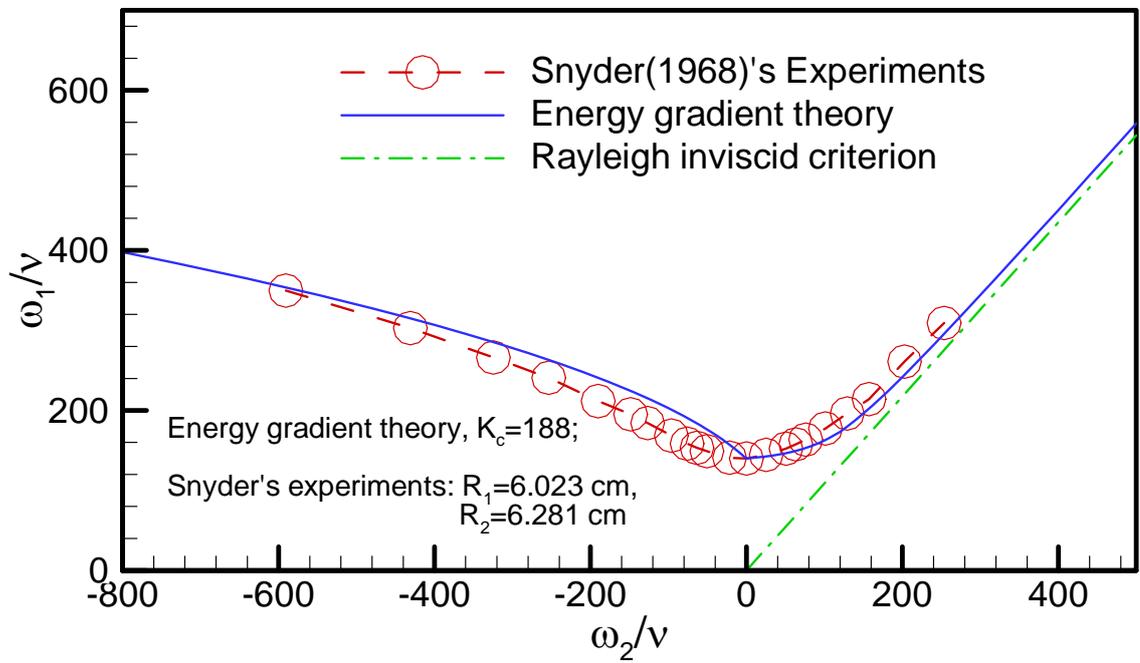

Fig.6 Comparison of the theory with the experimental data for the instability condition of Taylor-Couette flow (Snyder (1968)'s experiments, R1=6.023 cm, R2=6.281 cm). The data are taken from Table III in [11]. The relative gap width is $h/R_1$=0.0428.

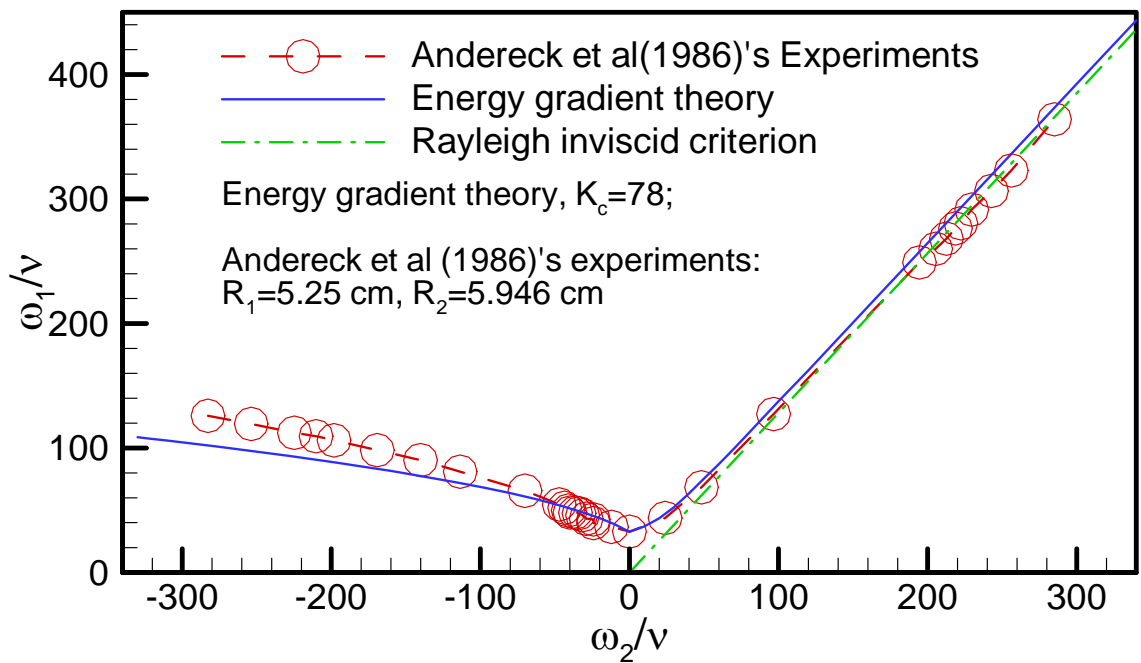

Fig.7 Comparison of the theory with the experimental data for the instability condition of Taylor-Couette flow (Andereck et al (1986)'s experiments, R1=5.25 cm, R2=5.946 cm). The data are taken from their Fig.2 and Fig.18 in [16]. The relative gap width is $h/R_1$=0.1326.

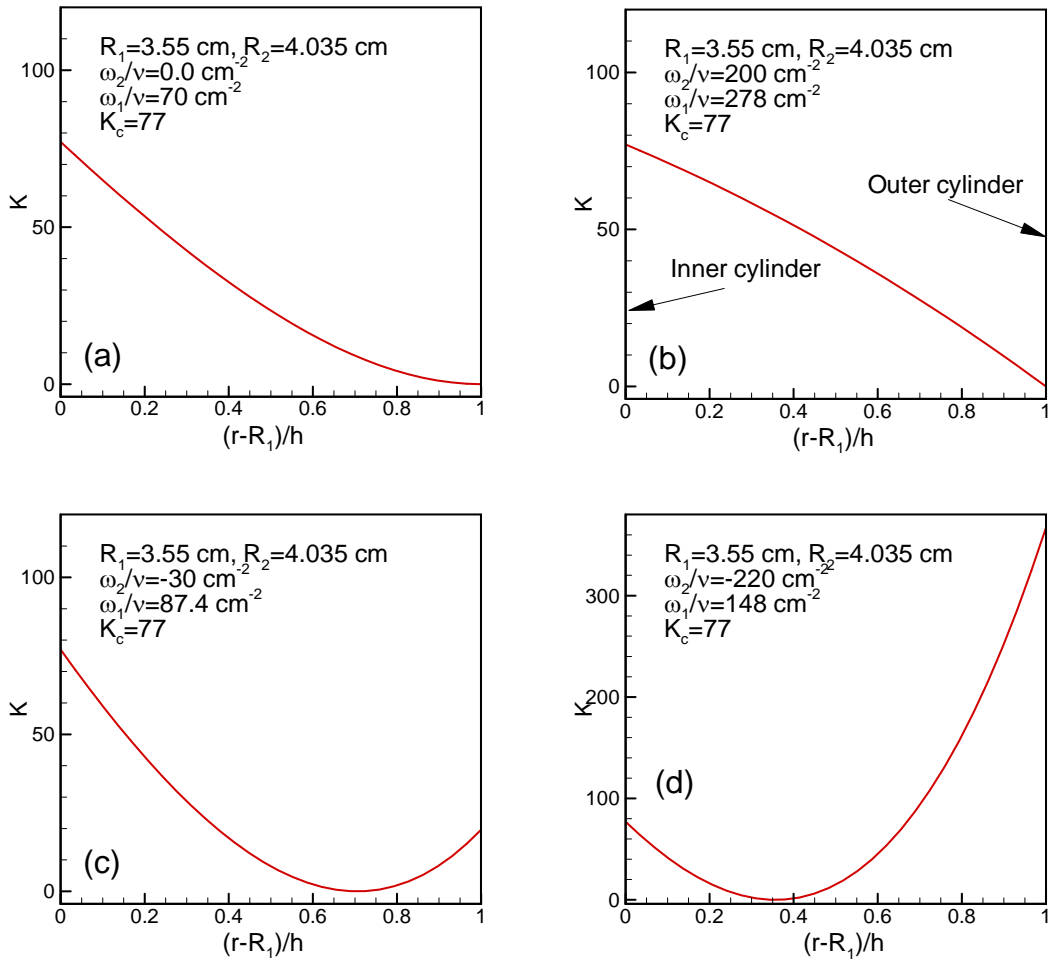

Fig.8 Distribution of K along the channel width at the critical condition $K_c=77$ corresponding to Fig.4. All the four data points are taken from the solid line calculated by the energy gradient theory in Fig.4. (a) The inner cylinder rotates while the outer cylinder is at rest; (b) Two cylinders rotate in same direction; (c) Two cylinders rotate in opposite directions and the speed of the outer cylinder is low. (d) Two cylinders rotate in opposite directions and the speed of the outer cylinder is high.

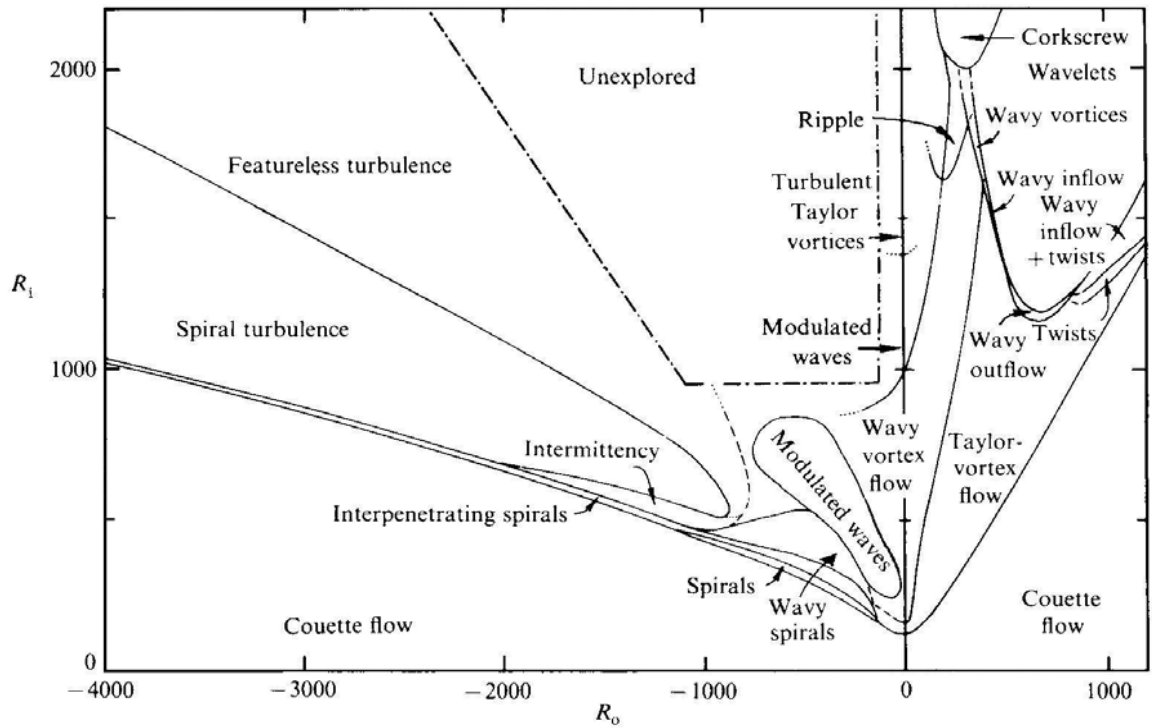

Fig.9 Regimes of the flow behaviour as identified by Andereck et al. [16]. The ordinate and abscissa are the Reynolds number based on the channel width and the circumferential velocities of the inner and outer cylinder, respectively (Used with permission by Cambridge University Press).

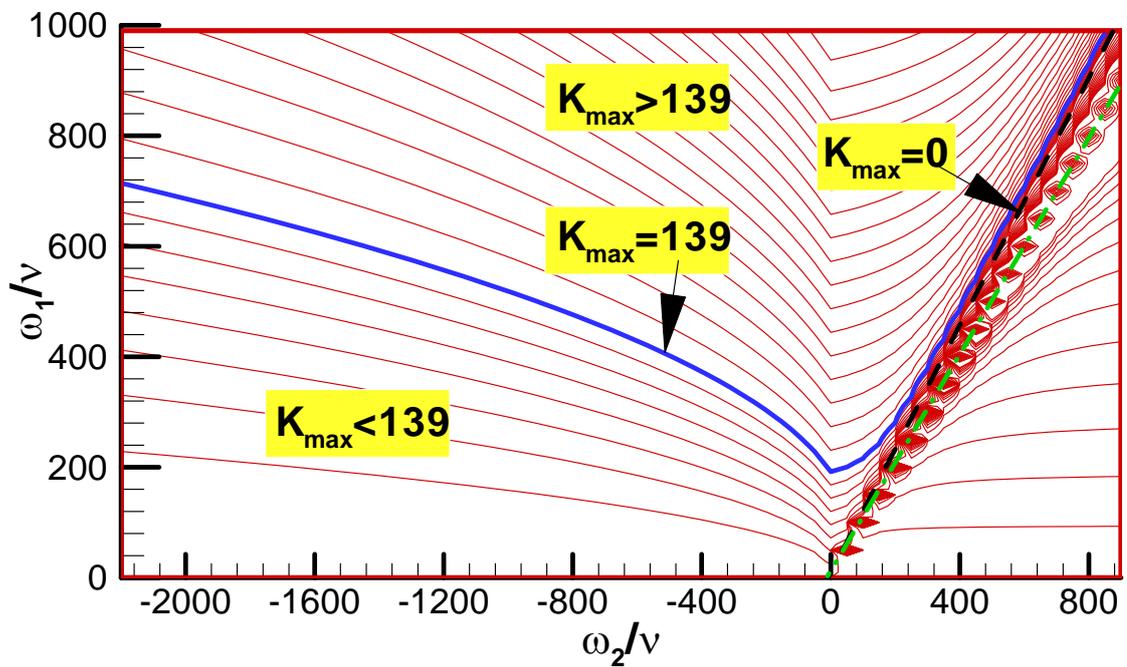

Fig.10 Isoline of $K_{max}$ along the inner cylinder in the plane of the rotating speeds of inner and outer cylinders (R1=3.80cm, R2=4.035 cm), corresponding to Fig.3. The critical value of $K_{max}$ is indicated in the figure by the thick blue line, which is calculated as shown in Figure 3.

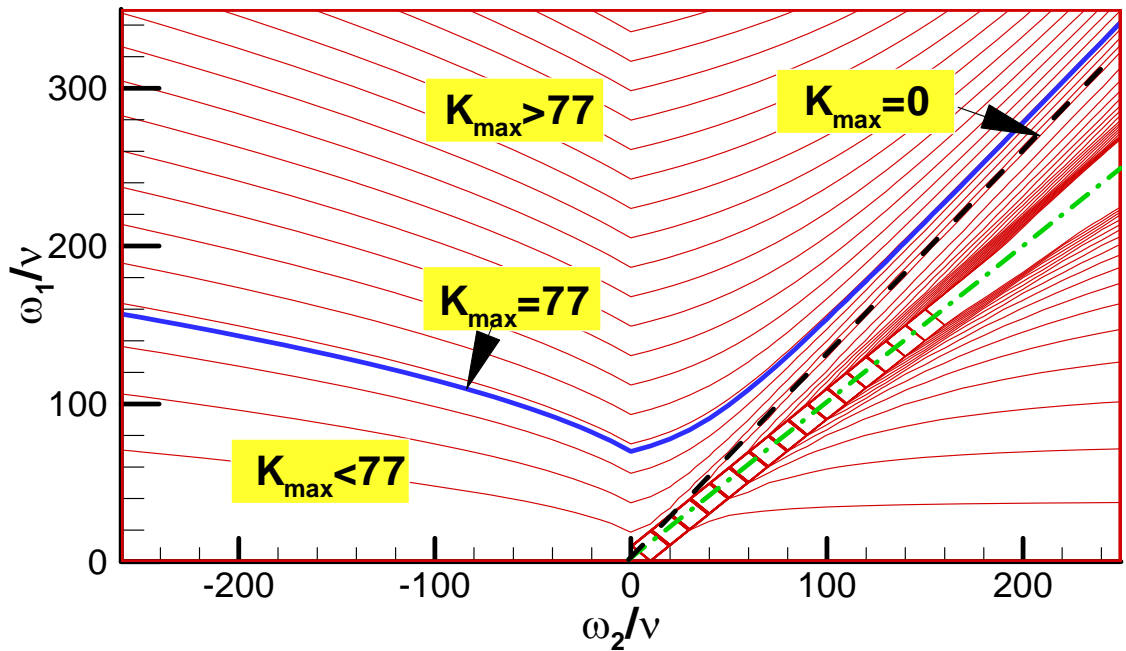

Fig.11 Isoline of $K_{max}$ along the inner cylinder in the plane of the rotating speeds of inner and outer cylinders (R1=3.55cm, R2=4.035 cm), corresponding to Fig.4. The critical value of $K_{max}$ is indicated in the figure by the thick blue line, which is calculated as shown in Figure 4.

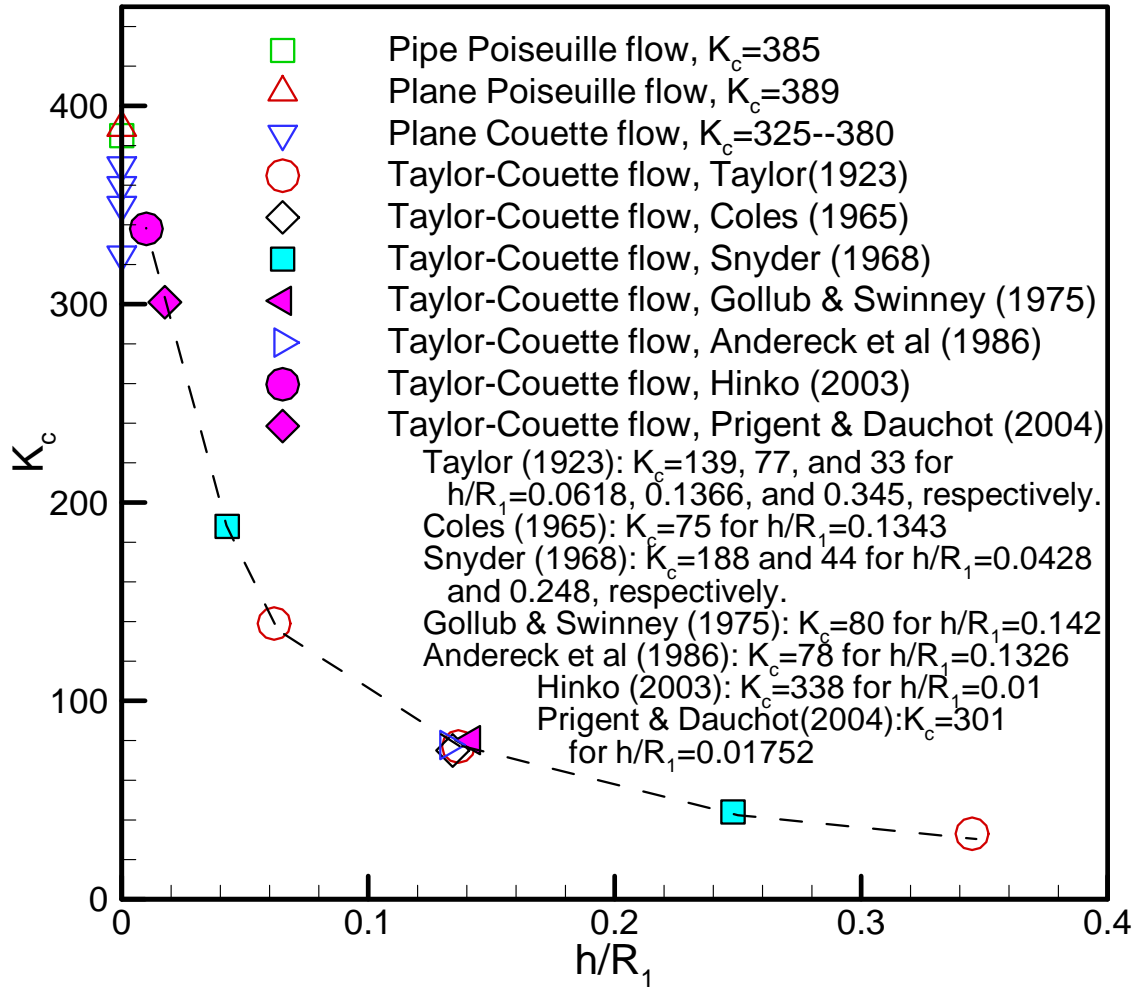

Fig.12 Critical value ($K_c$) of the energy gradient parameter $K_{max}$ versus parameter $h/R_1$ for Taylor-Couette flows. A dashed line to connect the data is drawn for visual convenience. The data for wall-bounded parallel flows (plane Poiseuille flow, pipe Poiseuille flow and plane Couette flow) are also shown, which are determined using the energy gradient theory in conjunction with the experimental data [21][22][23].